\documentclass[conference]{IEEEtran}
\IEEEoverridecommandlockouts
\usepackage{graphicx}
\usepackage{amsmath,bbm,epsfig,amssymb,amsfonts, amstext, verbatim,amsopn,cite,subfigure,multirow,multicol,lipsum}
\usepackage{balance}
\usepackage{booktabs,threeparttable}
\usepackage{amsfonts}
\usepackage{tikz}
\usepackage{enumerate} 
\allowdisplaybreaks
\usepackage{epstopdf} 
\usepackage{amsmath,bbm,amssymb,amsfonts,amstext,amsopn}

\usepackage{epsfig}
\usepackage{epstopdf}
\usepackage[process=auto]{pstool}
\usepackage{epsfig}
\usepackage{caption}
\usepackage{setspace}
\usepackage{stmaryrd}
\usepackage{multirow}
\usepackage{float}
\usepackage[process=auto]{pstool}
\usepackage{etoolbox}
\usepackage{algorithm}
\usepackage{algorithmic}
\usepackage{etoolbox}
\usepackage{multicol,lipsum}
\newcommand{\subparagraph}{}
\usepackage[compact]{titlesec}
\usepackage[english]{babel}
\usepackage[process=auto]{pstool}
\usepackage{amsthm}
\theoremstyle{remark}

\usepackage{etoolbox}
\usepackage{algorithm}
\usepackage[process=auto]{pstool}
\usepackage{epsfig}
\usepackage{caption}
\usepackage{algorithmic}
\usepackage{xcolor}
\usepackage{hyperref}
\usepackage{pifont}
\allowdisplaybreaks
\usepackage{amsmath,amssymb,amsfonts}
\usepackage{algorithmic}
\usepackage{graphicx}
\usepackage{textcomp}
\usepackage{bm}
\usepackage{amsmath,bbm,amssymb,amsfonts,amstext,amsopn}
\usepackage{amssymb}

\usepackage{amsthm}
\DeclareMathOperator{\Tr}{Tr}
\DeclareMathOperator{\Rank}{Rank}
\usepackage{cite}
\usepackage{balance}
\allowbreak
\usepackage{url}
\usepackage{epsfig}
\usepackage{setspace}
\usepackage{stmaryrd}
\usepackage{psfrag}	
\usepackage{lipsum,amsmath,multicol}
\usepackage{amsthm}
\usepackage{amssymb}
\usepackage{epstopdf}
\usepackage{amsthm}
\usepackage{cite}
\usepackage{balance}
\allowbreak
\usepackage{url}
\usepackage{epsfig}
\usepackage{setspace}
\usepackage{stmaryrd}
\usepackage{psfrag}	
\usepackage{multirow}
\usepackage{float}
\usepackage{amsthm}
\usepackage{float}
\usepackage{amsthm}

\usepackage[compact]{titlesec}
\usepackage[english]{babel}
\usepackage[process=auto]{pstool}
\usepackage{amsthm}

\theoremstyle{remark}
\usepackage{epstopdf}
\usepackage{caption}
\allowdisplaybreaks
\usepackage{algorithmic}
\usepackage{xcolor}
\usepackage{etoolbox}
\usepackage{algorithm}
\usepackage[process=auto]{pstool}
\usepackage{epsfig}

\begin{document}
\title{Codebook Based Two-Time Scale Resource Allocation Design for  IRS-Assisted \\ eMBB-URLLC Systems}
\author{ Walid R. Ghanem$^{*}$,
	Vahid Jamali$^{*}$, Malte Schellmann$^{\dagger}$, Hanwen Cao$^{\dagger}$, Joseph Eichinger$^{\dagger}$, and Robert Schober$^{*}$  \\
	$^{*}$Friedrich-Alexander-University Erlangen-Nuremberg, Erlangen, Germany\\ $^{\dagger}$Huawei Technologies German Research Center, Munich, Germany \quad \vspace{-0.35cm}}
\maketitle
\begin{abstract}	
This paper investigates the resource allocation algorithm design for wireless systems assisted by large intelligent reflecting surfaces (IRSs) with coexisting enhanced mobile broadband (eMBB) and ultra reliable low-latency communication (URLLC) users. We consider a two-time scale resource allocation scheme, whereby the base station's precoders are optimized in each mini-slot to adapt to newly arriving URLLC traffic, whereas the IRS phase shifts are reconfigured only in each time slot to avoid excessive base station-IRS signaling. To facilitate efficient resource allocation design for large IRSs, we employ a codebook-based optimization framework, where the IRS is divided into several tiles and the phase-shift elements of each tile are selected from a pre-defined codebook. The resource allocation algorithm design is formulated as an optimization problem for the maximization of the average sum data rate of the eMBB users over a time slot while guaranteeing the quality-of-service (QoS) of each URLLC user in each mini-slot. An iterative algorithm based on alternating optimization (AO) is proposed to find a high-quality suboptimal solution. As a case study, the proposed algorithm is applied in an industrial indoor environment modelled via the Quadriga channel simulator. Our simulation results show that the proposed algorithm design enables the coexistence of eMBB and URLLC users and yields large performance gains compared to three baseline schemes. Furthermore, our simulation results reveal that the proposed two-time scale resource allocation design incurs only a small performance loss compared to the case when the IRSs are optimized in each mini-slot.   	 
\end{abstract}
\section{Introduction}  \vspace{-0.2cm}
Enhanced mobile broadband (eMBB) and ultra reliable low-latency communications (URLLC) are two main service categories in the fifth-generation (5G) and beyond wireless networks \cite{chsejoint}. Specifically, eMBB applications require extremely high data rates, while URLLC services demand low latency and high reliability. The joint scheduling of eMBB and URLLC users is challenging due to their different quality-of-service (QoS) requirements. Several studies have investigated the joint scheduling of eMBB and URLLC traffic \cite{Anand}. However, meeting the eMBB and URLLC requirements may not be possible when the wireless channel conditions are unfavourable, e.g., due to blockages. A promising emerging technology to cope with this problem are intelligent reflecting surfaces (IRSs). An IRS comprises a set of passive elements which can reflect the incident signals to desired directions by applying appropriate phase shifts\cite{Quirs1}. By optimizing the IRS phase shifts, wireless channels can be customized, and virtual line-of-sight (LoS) links to the users can be established \cite{Quirs1}. Thus, IRSs can help enhance the data rates of eMBB users and increase the reliability and reduce the delay of URLLC users, especially when the users do not have a direct LoS to the base station (BS). 

Motivated by the ability of IRSs to configure and control the wireless propagation environment, IRS phase-shift optimization has been studied extensively in the literature  \cite{Quirs1,ofdmairs,alex1}. For example, resource allocation for IRS-aided orthogonal frequency division multiple access (OFDMA) systems was considered in \cite{ofdmairs}. In \cite{alex1}, robust and secure resource allocation algorithms for IRS-assisted systems were proposed.
  Existing works such as \cite{ofdmairs,alex1} focused mainly on eMBB traffic. A few recent studies also investigated IRS-assisted URLLC. The authors in \cite{ghanem6} propose to maximize the URLLC sum rate in IRS-OFDMA multi-cell systems. In \cite{ghanem6}, the BS serves only URLLC users. In contrast, in \cite{irsembb}, the coexistence of eMBB and URLLC users is studied for a single-antenna BS. However, modern communication systems are expected to employ multiple-antenna BSs. Moreover, the IRS phase shifts in \cite{irsembb} are updated in each mini-slot which introduces additional delays for the URLLC users due to the required additional signalling overhead between BS and IRS.     

Furthermore, the algorithms developed in \cite{Quirs1,ofdmairs,alex1,ghanem6,irsembb} aim to optimize all IRS elements individually which leads to high complexity and a huge signalling overhead particularly when the IRS is large \cite{marirsj}. To overcome these issues, the authors of \cite{marirsj} proposed to design a codebook of phase-shift configurations in an offline stage, and then select the best phase-shift configuration from the codebook in an online stage. In this case, the complexity of online optimization and the signalling overhead scale with the IRS codebook size and not with the number of reflecting elements. IRS codebook designs were reported in \cite{marirsj,quad}, and \cite{ghanem7code}.  Moreover, codebook based online optimization of IRS-assisted wireless systems was considered in \cite{marirsj,Dongfangirsscal}. Furthermore, recent studies \cite{two_time_scaleirs1,two_time_scaleirs2} have investigated two-time scale optimization of IRS systems by optimizing the IRS based on statistical channel state information (CSI).  However, the results in \cite{marirsj,Dongfangirsscal,two_time_scaleirs1,two_time_scaleirs2} are not applicable to eMBB-URLLC systems due to the different arrival times of the users. Thus, the efficient optimization of IRS-assisted  eMBB-URLLC systems is still an open problem.  

This paper makes the following main contributions:
\begin{itemize}
	\item We propose a novel resource allocation algorithm design for an eMBB-URLLC system assisted by large IRSs. The proposed resource allocation algorithm is formulated as a two-time scale resource allocation problem, whereby the precoders at the BS are optimized in each mini-slot to adapt to the newly arriving URLLC traffic, whereas the IRS phase shifts are reconfigured only in each time slot to avoid excessive BS-IRSs signaling. Then, the resource allocation algorithm design is formulated as a non-convex optimization problem for maximization of the average sum data rate of the eMBB users subject to QoS constraints for the URLLC users based on a low-complexity codebook based IRS design \cite{marirsj}.
	
	\item Finding the globally optimal solution entails a high computational complexity which is not desirable for URLLC real-time applications. Therefore, a suboptimal iterative algorithm based on alternating optimization (AO) is proposed to find a high-quality suboptimal solution.
	\item Our simulation results show that the proposed algorithm design facilitates the coexistence of eMBB and URLLC users and yields large performance gains compared to three baseline schemes. 
\end{itemize}

\textit{Notation}: In this paper, $\log(\cdot)$ is the logarithm with base 2. $\Tr{(\mathbf{A})}$ and $\Rank{(\mathbf{A})}$ denote the trace and the rank of matrix $\mathbf{A}$, respectively. $\mathbf{A}\succeq 0$  indicates that matrix $\mathbf{A}$ is positive semi-definite. $\mathbf{A}^{H}$ and ${\mathbf{A}}^{T}$ denote the Hermitian transpose and the transpose of matrix $\mathbf{A}$, respectively.  $\mathbb{C}$ is the set of complex numbers. $\mathbb{H}_{N}$  denotes the set of all $N \times N$ Hermitian matrices. $|\cdot|$ and  $\|\cdot\|$ refer to the absolute value of a complex scalar and the Euclidean vector norm, respectively. The circularly symmetric complex Gaussian distribution with mean $\mu$ and variance $\sigma^{2}$ is denoted by $\mathcal{CN}(\mu,\sigma^{2})$, and $\sim$ stands for ``distributed as". $\mathcal{E}\{\cdot\}$ denotes statistical expectation. $\nabla_{x}f(\mathbf{x})$ denotes the gradient vector of function $f(\mathbf{x})$ and its elements are the partial derivatives of $f(\mathbf{x})$. $|\mathcal{A}|$ is the cardinality of set $\mathcal{A}$.

\section{System and Signal Models}\vspace{-0.2cm}
In this section, we present the system and signal models for the considered IRS-assisted eMBB-URLLC system.
\subsection{System Model}
\begin{figure}[t]
	\centering
	\scalebox{0.65}{
		\pstool{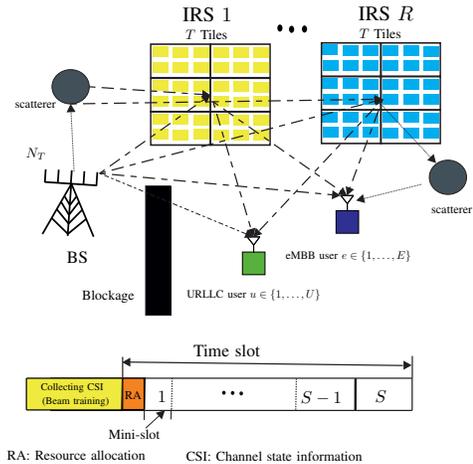}{
			\psfrag{BS}[c][c][1]{BS}
			\psfrag{sc}[c][c][0.7]{scatterer}
			\psfrag{nt}[c][c][0.8]{$N_{T}$ }
			\psfrag{u1}[c][c][0.65]{ URLLC user $u\in\{1,\dots,U\}$ }
			\psfrag{u4}[c][c][0.65]{eMBB user $e\in\{1,\dots,E\}$}
			\psfrag{B}[c][c][0.8]{Blockage}
			\psfrag{irs 1}[c][c][1.2]{IRS $1$}
			\psfrag{irs 2}[c][c][1.2]{IRS $R$}
			\psfrag{q}[c][c][0.8]{$T$ Tiles}
			\psfrag{C}[c][c][0.6]{Collecting CSI}
			\psfrag{C1}[c][c][0.6]{  (Beam training)}
			\psfrag{CT}[c][c][1]{Time slot}
			\psfrag{RA}[c][c][0.7]{RA}
			\psfrag{t1}[c][c][0.6]{Time slot $1$}
			\psfrag{tn}[c][c][0.6]{Time slot $N$}
			\psfrag{m}[c][c][1]{ $1$}
			\psfrag{sm}[c][c][1]{$S$}
			\psfrag{sm-1}[c][c][1]{$S-1$}
			\psfrag{ms}[c][c][0.8]{Mini-slot}
			\psfrag{ra}[c][c][0.8]{RA: Resource allocation \quad}
			\psfrag{csi}[c][c][0.8]{CSI: Channel state information}
	}}
	\caption{ $R$ IRSs assist a multiuser eMBB-URLLC system 
		comprising one BS, $K$ users, i.e., $E$ eMBB users and $U$ URLLC users.}\vspace{-0.4cm}
	\label{figs}
\end{figure}

We consider a single-cell downlink system, where a BS equipped with $N_{T}$ antennas serves $K$ single-antenna users, namely, $E$ eMBB and $U$ URLLC users, see Fig.~\ref{figs}, using the same frequency resource of $W$~Hz. To enhance the system performance, $R$ IRSs each comprising $T$ tiles and each tile containing $Q$ phase-shift elements are deployed to assist the BS in ensuring the QoS of the users. Similar to the 5G New
Radio (NR) standard \cite{3Gpp2a}, we assume that a resource frame
has a duration of $T_{f}$~\textrm{s} and consists of $N_{s}$ time slots of length $T_{s}$~\textrm{s}, and each time slot consists of $S$ mini-slots of duration $T_{ms}$~\textrm{s}. The resource allocation is performed per time slot to adapt to new incoming users. The eMBB users are admitted at the beginning of each time slot while the URLLC users may arrive within the mini-slots \cite{3Gpp2a}.

\textbf{Two-Time Scale Optimization}:
 Given the stringent delay requirements of URLLC traffic, an URLLC user should be served immediately in the next mini-slot upon its arrival to avoid queuing delays. Thus, the BS has to configure its precoders and the IRS phase shifts to support all possible incoming URLLC traffic scenarios. However, configuring the IRS phase shifts within the mini-slots adds additional delays and signaling  overhead which may not be affordable due to the URLLC delay constraint. To solve this problem, the IRS phase shifts are optimized once per time slot, while the precoders are adapted in each mini-slot. Since it is not a priori known which URLLC users arrive in a mini-slot, the proposed resource allocation scheme accounts for all $L=2^{U}$ possible combinations of active URLLC users.

\textbf{Tile and Codebook Based IRS Configuration}:
To facilitate low-complexity and scalable configuration of large IRSs, we adopt the tile and codebook based design framework from \cite{marirsj}. 
In particular, we optimize each tile in two stages, namely an offline design stage and an online optimization stage. In the offline design stage, we generate a reflection codebook by designing $M_{T}$ phase-shift configurations for the elements of each IRS tile \cite{marirsj,quad,ghanem7code}. In this paper, we adopt the existing codebook designs from  \cite{marirsj,quad,ghanem7code}, and focus our attention on the online optimization. We note that the phase-shift codebooks are designed to account for all possible channel realizations. However, for a given  realization of the channel, there are only a few phase-shift configurations in the codebook that lead to significant received power and are relevant for online optimization. Let $M$ denote the number of relevant phase-shift configurations for a given channel realization, where typically $M\ll M_{T}$ holds. These relevant configurations can be determined during beam training and do not change during the coherence time of the channel \cite{beamirs}. The focus of this paper is the joint online optimization of the phase-shift configurations of all tiles and the BS precoder.    
  
In order to facilitate the presentation, in the following, we use superscript $o \in \{e,u\}$ to denote eMBB and URLLC users, respectively. Moreover,  we consistently use $i$ and $j$ as the indices of eMBB and URLLC users, respectively, and index $k$ when we generally refer to both eMBB and URLLC users.
\subsection{Signal Model}
In each mini-slot, a subset of the URLLC users may be active. Let us define the set of all possible combinations of active users as $\mathcal{A}=\{\mathcal{U}_{l}, \forall l=[1,\dots ,2^{U}]\}=\{\emptyset,(1),(1,2), (1,3)\dots, (1,2,\dots,U)\}$, e.g., set $\mathcal{U}_{1}=\emptyset$ means that no URLLC user is active, while set $\mathcal{U}_{3}=(1,2)$ means that URLLC users $1$ and $2$ are active and $\mathcal{U}_{l}$ is the set that contains the indices of the $l$-th active URLLC user set.  
We employ linear transmit precoding at the BS, where each user is assigned a unique beamforming vector. Hence, the transmit signal vector at the BS in a mini-slot can be written as follows: \vspace{-0.2cm}
\begin{align}
\mathbf{x}_{l}=\sum_{i=1}^{E}\mathbf{w}^{e}_{i,l}s^{e}_{i}+\sum_{j \in \mathcal{{U}}_{l}}\mathbf{w}^{u}_{j,l}s^{u}_{j},
\end{align}
where $s^{o}_{k}\in \mathbb{C}$ and $\mathbf{w}^{o}_{k,l}\in \mathbb{C}^{N_{T} \times 1}$ are the transmit symbol and the beamforming vector of user $k$ of traffic type $o$ at the BS for active user set $\mathcal{U}_{l} \in \mathcal{A}$, respectively. Moreover, without loss of generality, we assume  $\mathcal{E}\{|s^{o}_{k}|^{2}\}=1, \forall k, \forall o$. Thus, the received signal at user $k$ for traffic type $o$ for the $l$-th active user set is given as follows:\vspace{-0.25cm}
\begin{align}
y^{o}_{k,l}=(\mathbf{v}_{k}^{oH}+\sum_{t=1}^{\overline{T}}\sum_{m=1}^{M}b_{t,m}\mathbf{h}_{t,m,k}^{oH})\mathbf{x}_{l}+z^{o}_{k},\forall o, k,l,
\end{align}
where $\mathbf{v}_{k}^{o} \in \mathbb{C}^{N_{t} \times 1}$ is the direct channel from the BS to user $k$ of type $o$,  $\overline{T}=RT$, and   $\mathbf{h}^{oH}_{t,m,k}=\mathbf{g}^{oH}_{t,k}\mathbf{\Phi}_{t,m}\mathbf{F}_{t}$ $\in \mathbb{C}^{1\times N_{t}}$ is the end-to-end channel for user $k$ of type $o$ generated by tile $t$ with phase-shift configuration  codeword $m$\footnote{We note that different tiles may belong to the same IRS or to different IRSs.}. Here, $\mathbf{g}^{o}_{t,k} \in \mathbb{C}^{Q \times 1}$ is the channel from IRS tile $t$ to user $k$, $\mathbf{\Phi}_{t,m} \in \mathbb{C}^{Q\times Q}$ is the diagonal phase-shift matrix corresponding to codeword $m$, $\mathbf{F}_{t} \in \mathbb{C}^{Q\times N_{t}}$ is the channel matrix from the BS to IRS tile $t$, and $b_{t,m}=\{0,1\}$ is a binary variable for the codeword selection at IRS tile $t$. $b_{t,m}=1$ means that tile $t$ employs codeword $m$, where $\sum_{m=1}^{M}b_{t,m}=1$, i.e., only one codeword can be chosen for tile $t$. $z^{o}_{k} \sim \mathcal{CN}(0, \sigma^{2})$ denotes additive white Gaussian noise with variance $\sigma^{2}$. 

The signal-to-interference-and-noise ratio (SINR) of URLLC user $j$ for the $l$-th active user set is given as follows:
\begin{align} \label{gammau} 
\gamma^{u}_{j,l}=\frac{f^{u}_{j,l}}{I^{u}_{j,l}+\sigma^{2}}, \forall j,l,
\end{align}
where
\begin{align}&\label{fu}
f^{u}_{j,l}=\bigg|(\mathbf{v}_{j}^{uH}+\sum_{t=1}^{\overline{T}}\sum_{m=1}^{M}b_{t,m}\mathbf{h}_{t,m,j}^{uH} )\mathbf{w}^{u}_{j,l}\bigg|^{2},\\&\label{lu}
I^{u}_{j,l}=\bigg|(\mathbf{v}_{j}^{uH}+\sum_{t=1}^{\overline{T}}\sum_{m=1}^{M}b_{t,m}\mathbf{h}_{t,m,j}^{uH} )\overline{\mathbf{w}}^{u}_{j,l}\bigg|^{2},
\end{align}
where $\overline{\mathbf{w}}^{u}_{j,l}=(\sum_{i=1}^{E}\mathbf{w}^{e}_{i,l}+\sum_{r \in \mathcal{U}_{l} \setminus {j}}\mathbf{w}^{u}_{r,l})$. The SINRs of the eMBB users are denoted by $\gamma^{e}_{i,l}, \forall i,l,$  and are defined similarly to (\ref{gammau}). Furthermore, to obtain a performance upper bound for IRS-assisted eMBB-URLLC systems, perfect CSI is assumed to be available at the BS for
resource allocation. \color{black}
\subsection{URLLC Achievable Rates}
For performance evaluation of finite blocklength transmission, the so-called normal approximation was developed \cite{thesis}. Mathematically, the maximum number of bits $B_{j,\textrm{req}}$ conveyed in a mini-slot in a packet of $n$ symbols with error probability $\epsilon_{j,\textrm{req}}$ can be approximated as follows \cite{thesis,chsejoint}
\begin{align}\label{C1}
B_{j,\textrm{req}}=n\log(1+\gamma_{j,\textrm{req}})-\log(\text{e})Q^{-1}(\epsilon_{j,\textrm{req}})\sqrt{n},
\end{align}
where $n=WT_{ms}$. 
Given a required number of bits, $B_{j,\textrm{req}}$, a required packet error probability, $\epsilon_{j,\textrm{req}}$, and a required number of symbols, $n$, the BS can compute the required SINR, i.e., $\gamma_{j,\textrm{req}}$ for the URLLC user. 
\section{Optimization Problem Formulation}
The proposed resource allocation design is calculated at the beginning of each time slot. The proposed algorithm optimizes the IRS phase shifts for the entire time slot, i.e., the $b_{t,m}, \forall t,m,$ are fixed during a time slot, while depending on URLLC traffic arrivals the precoders at the BS are adapted to the active users in each mini-slot, i.e., $\mathbf{w}^{o}_{k,l}, \forall o,k,l,$ depends on the active URLLC user set $\mathcal{U}_{l}$. Thus, we formulate an optimization problem for maximization of the average data rate of the eMBB users while guaranteeing the QoS for each URLLC user in each mini-slot. More specifically, the optimization problem is formulated as follows: 
\begin{align}\label{OP1a}
&\text{P}_{0}:\quad \underset  {\mathbf{w}^{o}_{k,l}, \mathbf{b}}{\text{maximize}}\quad\sum_{l=1}^{2^{U}}p_{l}\sum_{i=1}^{E}\log(1+\gamma^{e}_{i,l})\\& \text{s.t.} \; \mathrm{C1}: f^{u}_{j,l} \geq \gamma_{j,\textrm{req}}(I^{u}_{j,l}+\sigma^{2}),\forall j \in \mathcal{U}_{l}, \forall l, \nonumber \\& \;\;\quad \mathrm{C2}: \sum_{i=1}^{E}\|\mathbf{w}^{e}_{i,l}\|^{2}+\sum_{j \in \mathcal{U}_{l}}\|\mathbf{w}^{u}_{j,l}\|^{2} \leq P_{\text{max}}, \forall l,\nonumber \\&\;\;\quad \mathrm{C3}: \sum_{m=1}^{M}b_{t,m}=1, \forall t,	\nonumber 
\; \mathrm{C4}: b_{t,m} \in \{0,1\}, \forall t,m,	\nonumber  
\end{align}
where $p_{l}$ is the probability that active user set $\mathcal{U}_{l}$ occurs, i.e., $\sum_{l=1}^{2^{U}}p_{l}=1$, and $\mathbf{b}$ denotes the collection of optimization variables $b_{t,m}, \forall t,m$. Constraint $\mathrm{C1}$ ensures the QoS of each user in all active user sets $\mathcal{U}_{l}$. $\mathrm{C2}$ is the total power budget of the BS for each active user set or equivalently for each mini-slot. Constraints $\mathrm{C3}$ and $\mathrm{C4}$ are imposed since only one codeword from the codebook can be selected for each tile. The two-time scale optimization arises from the fact that the precoder $\mathbf{w}^{o}_{k,l}$ is optimized for each URLLC user set $\mathcal{U}_{l}$ that may occur in a mini-slot, while the IRS is optimized for the entire time-slot via variable $\mathbf{b}$ and does not change from one mini-slot to the next.  

Optimization problem (\ref{OP1a}) is a non-convex mixed-integer problem. The problem is difficult to solve since the optimization variables in the objective function and constraint $\mathrm{C1}$ are coupled, the SINR in the objective function and in $\mathrm{C1}$  has a non-convex structure, and the binary constraint $\mathrm{C4}$ is non-convex. Since, there is no systematic approach for solving general non-convex optimization problems, we propose a suboptimal AO based iterative algorithm for finding a stationary point for problem (\ref{OP1a}).
\section{Solution of The Optimization Problem}
In this section, we focus on finding a low-complexity suboptimal solution for problem (\ref{OP1a}) based on the AO approach, where we employ the Big-M formulation and successive convex approximation (SCA) techniques. The algorithm tackles the coupling of variables $\mathbf{w}^{o}_{k,l}$ and $\mathbf{b}$ by dividing problem (\ref{OP1a}) into two sub-problems, i.e., we alternatingly update $\mathbf{w}^{o}_{k,l}$ and  $\mathbf{b}$  while keeping the other variable fixed. The first sub-problem is given as follows: \vspace{-0.2cm}
\begin{align}\label{OPAO1}\hspace{-0.5cm}
&\text{P}_{1}:\quad \underset  {\mathbf{w}^{o}_{k,l}}{\text{maximize}}\quad\sum_{l=1}^{2^{U}}p_{l}\sum_{i=1}^{E}\log\big(1+\gamma^{e}_{i,l}(\mathbf{w}^{o}_{k,l},\mathbf{b}^{(a)})\big)\\& \text{s.t.} \; \overline{\mathrm{C1}}: \overline{f}^{u}_{j,l}(\mathbf{w}^{o}_{k,l},\mathbf{b}^{(a)}) \nonumber\\&\geq \gamma_{j,\textrm{req}} \big(\overline{I}^{u}_{j,l}(\mathbf{w}^{o}_{k,l},\mathbf{b}^{(a)})+\sigma^{2}\big),\forall j \in \mathcal{U}_{l}, \forall l, \mathrm{C2},\nonumber
\end{align}
where $\overline{f}^{u}_{j,l}(\mathbf{w}^{o}_{k,l},\mathbf{b}^{(a)})$ and $\overline{I}^{u}_{j,l}(\mathbf{w}^{o}_{k,l},\mathbf{b}^{(a)})$ are given by $f^{u}_{j,l}$ and $I^{u}_{j,l}$ for given $\mathbf{b}=\mathbf{b}^{(a)}$, respectively. $a$ is the AO iteration index. The second sub-problem is given by: \vspace{-0.2cm}
\begin{align}\label{OPAO2}
&\text{P}_{2}:\quad \underset  { \mathbf{b}}{\text{maximize}}\quad\sum_{l=1}^{2^{U}}p_{l}\sum_{i=1}^{E}\log(1+\gamma^{e}_{i,l}(\mathbf{w}^{o(a)}_{k,l},\mathbf{b}))\\& \text{s.t.} \;\widetilde{ \mathrm{C1}}: \widetilde{f}^{u}_{j,l}(\mathbf{w}^{o(a)}_{k,l},\mathbf{b})\nonumber\\& \geq \gamma_{j,\textrm{req}} \big(\widetilde{I}^{u}_{j,l}\big(\mathbf{w}^{o(a)}_{k,l},\mathbf{b})+\sigma^{2}\big) ,\forall j \in \mathcal{{U}}_{l}, \forall l, \nonumber  \mathrm{C3},\mathrm{C4},	\nonumber  
\end{align}
where $\widetilde{f}^{u}_{j,l}(\mathbf{w}^{o(a)}_{k,l},\mathbf{b})$ and $\widetilde{I}^{u}_{j,l}(\mathbf{w}^{o(a)}_{k,l},\mathbf{b})$ are given by $f^{u}_{j,l}$ and $I^{u}_{j,l}$ for given $\mathbf{w}^{o}_{k,l}=\mathbf{w}^{o(a)}_{k,l}$, respectively. In the following sub-sections, we propose two suboptimal algorithms for solving sub-problems (\ref{OPAO1}) and (\ref{OPAO2}). 
\subsection{Sub-problem 1 (Optimization of BS Precoders  $\mathbf{w}^{o}_{k,l}$)}
Optimization problem (\ref{OPAO1}) can be solved using a modified version of the suboptimal algorithm proposed in \cite{ghanem6}. In particular, the algorithm in \cite{ghanem6} is modified to account for the eMBB traffic and the different active user sets.  Due to space limitations, we cannot review this algorithm here and refer interested readers to \cite{ghanem6}. In the following, this modified algorithm is referred to \textbf{Algorithm} 1.
\subsection{Sub-problem 2 (Optimization of Codeword Selection $\mathbf{b}$)}
In this sub-section, we optimize the codeword selection for a given beamformer $\mathbf{w}_{k,l}^{o(a)}$. For a given  $\mathbf{w}_{k,l}^{o(a)}$, the SINR of URLLC user $j$ can be rewritten as follows \cite{marirsj}:\vspace{-0.2cm}
\begin{align} \label{gam2} \small
\widetilde{{\gamma}}^{u}_{j,l}=\frac{\widetilde{{f}}^{u}_{j,l}}{\widetilde{{I}}^{u}_{j,l}+\sigma^{2}},
\end{align}
where $\widetilde{{f}}^{u}_{j,l}$ and $\widetilde{{I}}^{u}_{j,l}$ are given as follows: 
\begin{align}& \label{I2a1} 
\widetilde{{f}}^{u}_{j,l}=\sum_{t=1}^{\overline{T}}\sum_{m=1}^{M}\sum_{p=1}^{\overline{T}}\sum_{q=1}^{M}b_{t,m}b_{p,q}\overline{\mathbf{h}}_{t,m,j}^{uH}\mathbf{w}^{u(a)}_{j,l}\mathbf{w}^{u(a)H}_{j,l}\overline{\mathbf{h}}^{u}_{p,q,j},\\& \label{I2b2} 
\widetilde{{I}}^{u}_{j,l}=\sum_{t=1}^{\overline{T}}\sum_{m=1}^{M}\sum_{p=1}^{\overline{T}}\sum_{q=1}^{M}b_{t,m}b_{p,q}\overline{\mathbf{h}}_{t,m,j}^{uH}\overline{\mathbf{w}}^{u(a)}_{j,l}\overline{\mathbf{w}}^{u(a)H}_{j,l}\overline{\mathbf{h}}^{u}_{p,q,j},
\end{align} 
where $\overline{\mathbf{h}}^{u}_{1,m,j}=\mathbf{h}^{u}_{1,m,j}+\mathbf{v}^{u}_{j}$ and $\overline{\mathbf{h}}^{u}_{t,m,j}={\mathbf{h}}^{u}_{t,m,j}, \forall t=\{2,\dots,\overline{T}\}$, i.e., the direct link is only added once to  the entire end-to-end channel for user $j$. \color{black} Since both $b_{t,m}$ and $b_{p,q}$ are binary variables, we apply the
big-M formulation\cite[Section~2.3]{Leemixed} to linearize the product terms $b_{t,m}b_{p,q}$. We first define new auxiliary optimization variables $\beta_{t,m,p,q}=b_{t,m}b_{p,q},  \forall t,p,\forall m,q,$ and decompose the product using the following linear convex constraints \cite{Leemixed}:\vspace{-0.25cm}
\begin{align}& 
\mathrm{C5a}: 0 \leq \beta_{t,m,p,q} \leq 1,  \mathrm{C5b}: \beta_{t,m,p,q} \leq b_{t,m},\\&
\mathrm{C5c}: \beta_{t,m,p,q}\leq b_{p,q},   \mathrm{C5d}:  \beta_{t,m,p,q} \geq b_{t,m}+b_{p,q}-1.
\end{align}
The binary constraint 
$\mathrm{C4}$ in problem (\ref{OPAO2}) is intrinsically non-convex. For handling this constraint, we rewrite it equivalently as follows:
\begin{align}&
\mathrm{C4a}:\sum_{t=1}^{\overline{T}}\sum_{m=1}^{M}b_{t,m}-b_{t,m}^{2} \leq 0, 
\mathrm{C4b}:0 \leq b_{t,m}\leq 1.
\end{align}
Constraint $\mathrm{C4a}$ is a difference of convex (D.C.) functions and we use the Taylor series approximation (TSA) to approximate the non-convex constraint as the following convex constraint \cite{Dongfangirsscal}: 
\begin{align}
\widetilde{\mathrm{C4a}}:\sum_{t=1}^{\overline{T}}\sum_{m=1}^{M}b_{t,m}-b^{2(i_{2})}_{t,m}-2b^{(i_{2})}_{t,m}(b_{t,m}-b^{(i_{2})}_{t,m}) \leq 0,
\end{align}
where $i_{2}$ is the SCA iteration index of \textbf{Algorithm} 2.
The objective function of problem (\ref{OPAO2}) is not convex with respect to $\beta_{t,m,p,q}$. To tackle this issue, we define new auxiliary variables $\widetilde{\chi}^{e}_{i,l}$ and $\widetilde{d}^{e}_{i,l}$ and rewrite optimization problem (\ref{OPAO2}) equivalently as follows:
\begin{align}\label{OPAO2a}\hspace{-1cm}
&\quad \underset  { \mathbf{b}, \boldsymbol{\beta},\widetilde{\chi}^{e}_{i,l},\widetilde{d}^{e}_{i,l}}{\text{maximize}}\quad\sum_{l=1}^{2^{U}}p_{l}\sum_{i=1}^{E}\log(1+\widetilde{\chi}^{e}_{i,l})\\& \text{s.t.} \; \widetilde{ \mathrm{C1}}: \widetilde{f}^{u}_{j,l}\geq \gamma_{j,\textrm{req}} (\widetilde{I}^{u}_{j,l}+\sigma^{2}),\forall j \in \mathcal{U}_{l}, \forall l, \nonumber \\&\quad\;\; \mathrm{C3},\widetilde{\mathrm{C4a}}, \mathrm{C4b},\mathrm{C5a-C5d},	\nonumber \\&   \quad\;\; \widetilde{\mathrm{C7}}:\widetilde{\chi}^{e}_{i,l}\widetilde{d}^{e}_{i,l}\leq  \widetilde{f}^{e}_{i,l},\forall i,l,  \widetilde{\mathrm{C8}}:\widetilde{d}^{e}_{i,l} \geq \sigma^{2}+\widetilde{I}^{e}_{i,l},\forall i,l. \nonumber
\end{align}
Constraint $\widetilde{\mathrm{C7}}$ can be convexified using TSA as follows:
\begin{align}&
\widetilde{\widetilde{\mathrm{C7}}}: 0.5({\widetilde{\chi}}^{e}_{i,l}+\widetilde{d}^{e}_{i,l})^{2}-0.5({\widetilde{\chi}}^{e(i_{2})}_{i,l})^{2}-{\widetilde{\chi}}^{e(i_{2})}_{i,l}({\widetilde{\chi}}^{e}_{i,l}-{\widetilde{\chi}}^{e(i_{2})}_{i,l})
\nonumber\\&-0.5({\widetilde{d}_{i,l}}^{e(i_{2})})^{2}-\widetilde{d}^{e(i_{2})}_{i,l}({\widetilde{d}}^{e}_{i,l}-\widetilde{d}^{e(i_{2})}_{i,l})\leq  \widetilde{f}^{e}_{i,l}.
\end{align}
By substituting constraint $\widetilde{\widetilde{\mathrm{C7}}}$ with $\widetilde{\mathrm{C7}}$, optimization problem (\ref{OPAO2}) is approximated as the following convex optimization problem: 
\begin{align}\label{OPAO2b}
&\quad \underset  { \mathbf{b}, \boldsymbol{\beta},\widetilde{\chi}^{e}_{i,l},\widetilde{d}^{e}_{i,l}}{\text{maximize}}\quad\sum_{l=1}^{2^{U}}p_{l}\sum_{i=1}^{E}\log(1+\widetilde{\chi}^{e}_{i,l})\\& \text{s.t.} \; \widetilde{ \mathrm{C1}},  \mathrm{C3},\widetilde{\mathrm{C4a}}, \mathrm{C4b},\mathrm{C5a-C5d}, \widetilde{\widetilde{\mathrm{C7}}}, \widetilde{\mathrm{C8}}. \nonumber
\end{align}
 Problem (\ref{OPAO2b}) is a convex problem that can be solved by standard convex solvers, e.g., CVX \cite{cvx}. \textbf{Algorithm} \ref{alg2} summarizes
the main steps for solving (\ref{OPAO2}) in an iterative manner, where the
solution of (\ref{OPAO2b}) in iteration $i_{2}$ is used as the initial point for
the next iteration $i_{2}+1$. Moreover, since problem (\ref{OPAO2}) is reformulated as
a D.C. problem and TSA is used to
convexify the problem, according to \cite{fastglobal}, \textbf{Algorithm} 2 produces a sequence
of improved feasible solutions until convergence to a local
optimum point of problem (\ref{OPAO2}) in polynomial time.  
\subsection{Convergence and Complexity of AO \textbf{Algorithm}}
The overall AO based algorithm is summarized in
\textbf{Algorithm} 3. The objective function in (\ref{OPAO1}) is monotonically non-decreasing in each iteration of \textbf{Algorithm} 1. Moreover, the objective function in (\ref{OPAO2}) is monotonically non-decreasing in each iteration of \textbf{Algorithm} 2. Thus, \textbf{Algorithm} 3 produces a non-decreasing sequence of optimization values until convergence to of a stationary point of problem (\ref{OP1a}), see \cite{aocon} for more details. 

The complexity order of \textbf{Algorithm} 3 depends on the complexity order of \textbf{Algorithms} 1 and 2. The complexity order of \textbf{Algorithms} 1 \cite{polik} is given by  $O\big(I_{1,\text{max}}((E+U)L+2EL)^{4}(2EL+(E+U)L+L(U+1))\big)$\cite{polik}, where $I_{1,\text{max}}$ is the required number of iterations of \textbf{Algorithm} 1. Similarly, the approximated overall complexity of \textbf{Algorithm} 2 is of order $O\big(I_{2,\text{max}}(\bar{T}M+\bar{T}^{2}M^{2}+2EL)^{4}(UL+\bar{T}M+\bar{T}+4\bar{T}^{2}M^{2}+2EL)\big)$\cite{polik}. 
	\begin{algorithm}[t] \setcounter{algorithm}{1}
	\begin{algorithmic}[1]
		\STATE {Initialize:}  Set initial points $\mathbf{w}_{k,l}^{o(a)}$, $\mathbf{b}^{(a)}$, and calculate $\boldsymbol{\beta}^{(a)}$. Set iteration index $i_{2}=1$, maximum number of iterations $I_{2,\text{max}}$. \\
		\STATE \textbf{Repeat}\\
		\STATE Use CVX for solving the approximated convex problem (\ref{OPAO2b}) for given $\mathbf{w}_{k,l}^{o(a)}$, $\mathbf{b}^{(i_{2})}$, and $\boldsymbol{\beta}^{(i_{2})}$ and store the intermediate solutions   $\mathbf{b}$ and $\boldsymbol{\beta}$.\\
		\STATE Set ${i_{2}}={i_{2}}+1$ and update $\mathbf{b}^{(i_{2})}=\mathbf{b}$ and $\boldsymbol{\beta}^{(i_{2})}=\boldsymbol{\beta}.$ \\
		\STATE \textbf{Until} convergence or $i_{2}=I_{2,\text{max}}$.\\		
		\STATE {Output:} $\mathbf{b}^{*}=\mathbf{b}^{(i_{2})}$ 
	\end{algorithmic}
	\caption{SCA-based Iterative Codewords Selection}
	\label{alg2}
\end{algorithm}
	\begin{algorithm}[t]
	\begin{algorithmic}[1]
		\STATE {Initialize:} Maximum number of iterations $A_{\text{max}}$, and iteration index $a$.  \\
		\STATE \textbf{repeat}\{\text{Main loop}\}\\
		\STATE Obtain ${\mathbf{w}}_{k,l}^{o(a+1)}$ using \textbf{Algorithm} 1 for given ${\mathbf{w}}_{k,l}^{o(a)}$, $\mathbf{b}^{(a)}$, and $\boldsymbol{\beta}^{(a)}$,  and store the intermediate solutions   ${\mathbf{w}}^{o(a+1)}_{k,l}$\\
		\STATE Obtain $\mathbf{b}^{(a+1)}$ using \textbf{Algorithm} 2 for given ${\mathbf{w}}_{k,l}^{o(a+1)}$, $\mathbf{b}^{(a)}$, and $\boldsymbol{\beta}^{(a)}$, and store the intermediate solutions   $\mathbf{b}^{(a+1)}$ and $\boldsymbol{\beta}^{(a+1)}$\\
		\STATE Set ${a}={a}+1$  
		\STATE \textbf{Until} convergence or $a=A_{\text{max}}$.\\		
		\STATE {Output:} $\mathbf{w}^{o*}_{k,l}=\mathbf{w}^{o(a)}_{k,l}$, $\forall o,k,l,$ 
		$\mathbf{b}^{*}=\mathbf{b}^{(a)}$.
	\end{algorithmic}
	\caption{Alternating Optimization Algorithm}
	\label{alg}
\end{algorithm}

\section{Performance Evaluation}
As a case study, in our simulations, we consider an industrial indoor factory environment. The IRSs are mounted at the walls of the factory to assist the BS in providing virtual LoS links to the users. We consider uniform rectangular arrays at the BS and the IRSs. The BS and two IRSs are located at $(-50,0,2)~\textrm{m}$, $(-30,30,6)~\textrm{m}$, and $(-30,-30,6)~\textrm{m}$, respectively. The users' height is $1.5$~\textrm{m}. The channel coefficients are generated using the QuaDRiGa channel simulator \cite{quadriga} for industrial indoor factory scenarios as specified in 3GPP 38.901 \cite{3Gpp38901}. The adopted path loss model is given in Table 7.4.1-1 in 3GPP 38.901 \cite{3Gpp38901}. We assume LoS links exist between the BS and the IRSs and between the IRSs and the users. We assume that the users are located in blocked areas of the factory. Thus, a shadowing attenuation of $h_{d}$~dB is applied to their direct links to the BS. The simulation parameters listed in Table \ref{tab1a} are employed, unless specified otherwise. Following a similar approach as in \cite{marirsj,ghanem7code}, for all tiles, we generate identical reflection codebooks
with $M_{T}$ codewords and a wavefront phase codebook of size $|\mathcal{B}_{0}|$\footnote{The overall codebook designed in the offline stage is the product of a reflection codebook and a wavefront phase codebook \cite{marirsj}. In particular, the reflection codebook enables the tile to reflect an incident signal with desired elevation and azimuth angles while the wavefront phase codebook facilitates the constructive or destructive combination of the signals that arrive from different tiles at the receivers.}. Then, we adjust the size of the reflection codebook by selecting the $M_{s}$ codewords which yield the largest received powers for each user during the beam training process. In particular, we calculate
the Euclidean norm of the strength of $\mathbf{h}^{o}_{t,m,k}$ to determine the
$M_{s}$ channel vectors for each user $k$. As a result, in our simulations, the number of codewords used for online optimization is given by $M = |\mathcal{B}_{0}|KM_{s}$.
\begin{table}[t]
	\centering
	\caption{Simulation parameters} 
	\renewcommand{\arraystretch}{1.4}
	\scalebox{0.42}{%
		\begin{tabular}{|c||c||c||c|}
			\hline
			 Carrier center frequency, $f_{c}$ & 3.75 GHz &			 Packet error probability, $\epsilon_{j,\textrm{req}}$  &   $10^{-6}$ \\ \hline
			Total bandwidth, $W$ & 1.2 MHz &			Number of scatters for each link &  5 \\ \hline
			Time-slot duration, $T_{s}$  & 0.5~\textrm{ms} &Number of antennas at the, BS $N_{T}$  &  6 \\ \hline
			Mini-slot duration, 	$T_{ms}$  & 70~$\mu$\textrm{s} & 	Noise power density, $N_{o}$  & -174 dBm/Hz \\ \hline
			Number of IRSs, $R$ &  2 & IRS elements per tile, $Q$ &  144  \\  \hline   
		 Number of tiles per IRS, 	$T$ &  4 & Codebook size, $M_{T}$  & $144$ \\  \hline    
		Spacing between BS antennas or IRS elements, $d_{y}=d_{z}$  &  $\frac{\lambda}{2}$ & 	Probability of each active user set, $p_{l}$&  $\frac{1}{L}$  \\  \hline  
			Number of iterations for \textbf{Algorithms} 1 and 2, $I_{1,\text{max}}$, $I_{2,\text{max}}$&  25 & 			Number of iterations for \textbf{Algorithm} 3, $A_{\text{max}}$&  25  \\  \hline    
			Size of wavefront codebook, \cite{marirsj} $|\mathcal{B}_{0}|$&  3 & 			 Number of selected codewords for each user, $M_{\text{s}}$ &  2 \\  \hline  
	\end{tabular}}
	\label{tab1a}\vspace{-0.45cm}
\end{table}
\subsection{Upper Bound and Baseline Schemes}
\par We compare the performance of the proposed resource allocation algorithm with the following upper bound and baseline schemes:
\begin{itemize}
	\setlength{\itemsep}{1pt}
		\item {\textbf{Upper bound}}: To obtain a performance upper bound, we optimize both the beamforming at the BS and the IRSs in mini-slot for each active user set. The corresponding optimization problem is solved using a modified version of \textbf{Algorithm} \ref{alg}.
\item {\textbf{Baseline scheme 1:}} In this scheme, we randomly select a codeword for each tile from the $M$ available codewords and optimize the beamforming at the BS for each active user set.
	\item {\textbf{Baseline scheme 2:}} In this scheme, we adopt random phase shifts for the IRSs elements and optimize the beamforming at the BS for each active user set.
	\item {\textbf{Baseline scheme 3:}} In this scheme, we remove the IRSs from the system model and optimize the beamforming at the BS in each mini-slot.
\end{itemize}
  \subsection{Simulation Results} 
\begin{figure}[t!]
	\centering
	\scalebox{0.5}{
		\psfragfig{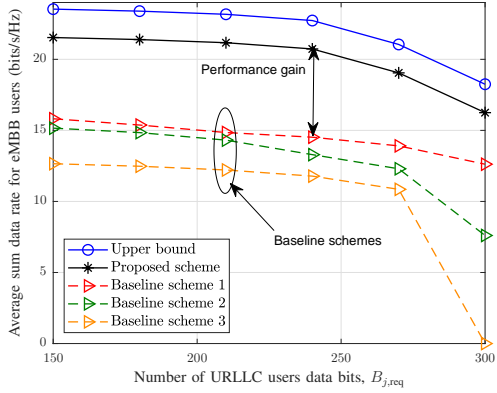}}
	\caption{Average sum data rate of eMBB users versus the minimum required number of URLLC user data bits, $B_{j,\text{req}}$. Number of eMBB users $E=2$, number of URLLC users $U=2$, maximum BS transmit power $P_{\text{max}}=28$~dB, and $h_{d}=25$~dB.}
	\label{Fig_s1}\vspace{-0.65cm}
\end{figure}

Fig.~\ref{Fig_s1} shows the average sum data rate of the eMBB users versus the minimum required number of information bits transmitted  to each URLLC user, $B_{j,\text{req}}$, for different resource allocation schemes. 
 As can be seen from Fig.~\ref{Fig_s1}, the proposed scheme achieves a large performance gain compared to baseline schemes 1, 2, and 3, thanks to the optimization of the IRSs. In particular, for baseline scheme 3, there are no IRSs in the system, and thus, the average sum data rate for the eMBB users is limited by the poor channel conditions between the users and the BS. Moreover, the proposed scheme attains large performance gains with respect to baseline schemes 1 and 2. Baseline scheme 2 applies random phase shifts and cannot exploit the benefits of passive IRS beamforming. In contrast, baseline scheme 1 applies random codewords from the $M$ available codewords, and thus, the reflected signals from different tiles may be destructively combined at the receiver. However, baseline scheme 1 attains a higher average sum data rate compared to baseline scheme 2, since the $M$ available codewords were properly preselected to steer the beam reflected by each tile towards the users, cf. Fig.~\ref{figs}. 
   
\begin{figure}[t!]
	\centering
	\scalebox{0.5}{
		\psfragfig{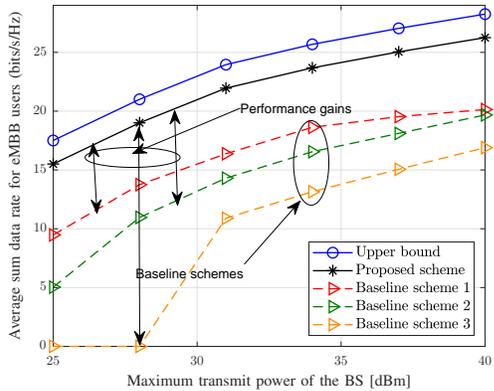}}
	\caption{Average sum data rate of eMBB users versus maximum BS transmit power. $B_{j,\text{req}}=180$~bits, number of eMBB users $E=2$, number of URLLC users $U=2$, and $h_{d}=30$~dB.}
	\label{Fig_s2}\vspace{-0.65cm}
\end{figure}

Fig.~\ref{Fig_s2} shows the average sum data rate of the eMBB users versus the maximum transmit power of the BS, $P_{\text{max}}$, for different resource allocation schemes. The proposed scheme attains large performance gains compared to the baseline schemes across the entire range of considered transmit powers. This is attributed to the optimization of the IRS phase shifts, which can successfully combat the poor channel conditions to facilitate high data rates for the eMBB users and ensure the QoS constraints of the URLLC users.

 Furthermore, from Figs.~\ref{Fig_s1} and \ref{Fig_s2}, we observe that the performance loss of the proposed scheme compared to the upper bound is relatively small. Recall that, for the upper bound, the IRSs are optimized in each mini-slot after the URLLC user activity is known. This leads to a large signaling overhead and long delays for the URLLC users.
 
\section{Conclusions}	
In this paper, we have studied the resource allocation algorithm design for wireless systems assisted by large IRSs with coexisting eMBB and URLLC users. We have proposed a two-time scale resource allocation scheme and adopted a codebook-based optimization framework. The resource allocation algorithm design was formulated as an optimization problem for maximization of the average data rate of the eMBB users over a time slot while guaranteeing the QoS of each URLLC user in each mini-slot. An iterative algorithm based on AO was developed to find a high-quality suboptimal solution. Our simulations have confirmed that the proposed algorithm design facilitates the coexistence of eMBB and URLLC users and yields large performance gains compared to three baseline schemes. 
\bibliography{ref}  

\begin{thebibliography}{10}
\providecommand{\url}[1]{#1}
\csname url@samestyle\endcsname
\providecommand{\newblock}{\relax}
\providecommand{\bibinfo}[2]{#2}
\providecommand{\BIBentrySTDinterwordspacing}{\spaceskip=0pt\relax}
\providecommand{\BIBentryALTinterwordstretchfactor}{4}
\providecommand{\BIBentryALTinterwordspacing}{\spaceskip=\fontdimen2\font plus
\BIBentryALTinterwordstretchfactor\fontdimen3\font minus
  \fontdimen4\font\relax}
\providecommand{\BIBforeignlanguage}[2]{{%
\expandafter\ifx\csname l@#1\endcsname\relax
\typeout{** WARNING: IEEEtran.bst: No hyphenation pattern has been}%
\typeout{** loaded for the language `#1'. Using the pattern for}%
\typeout{** the default language instead.}%
\else
\language=\csname l@#1\endcsname
\fi
#2}}
\providecommand{\BIBdecl}{\relax}
\BIBdecl

\bibitem{chsejoint}
C.~She, C.~Yang, and T.~Q.~S. Quek, ``Joint uplink and downlink resource
  configuration for ultra-reliable and low-latency communications,''
  \emph{{IEEE} Trans. Commun}, vol.~66, no.~5, pp. 2266--2280, May 2018.

\bibitem{Anand}
A.~Anand, G.~De~Veciana, and S.~Shakkottai, ``Joint scheduling of {URLLC} and
  {eMBB} traffic in {5G} wireless networks,'' in \emph{Proc. IEEE Conf. on
  Computer Commun.}, 2018, pp. 1970--1978.

\bibitem{Quirs1}
Q.~{Wu} and R.~{Zhang}, ``Intelligent reflecting surface enhanced wireless
  network via joint active and passive beamforming,'' \emph{IEEE Trans. Wirel.
  Commun.}, vol.~18, no.~11, pp. 5394--5409, Aug. 2019.

\bibitem{ofdmairs}
Y.~{Yang}, S.~{Zhang}, and R.~{Zhang}, ``{IRS}-enhanced {OFDMA}: Joint resource
  allocation and passive beamforming optimization,'' \emph{IEEE Wireless
  Commun. Lett.}, vol.~9, no.~6, pp. 760--764, Jan. 2020.

\bibitem{alex1}
X.~{Yu}, D.~{Xu}, Y.~{Sun}, D.~W.~K. {Ng}, and R.~{Schober}, ``Robust and
  secure wireless communications via intelligent reflecting surfaces,''
  \emph{IEEE J. Sel. Areas Commun.}, vol.~38, no.~11, pp. 2637 -- 2652, July
  2020.

\bibitem{ghanem6}
W.~R. Ghanem, V.~Jamali, and R.~Schober, ``Joint beamforming and phase shift
  optimization for multicell {IRS-aided OFDMA-URLLC} systems,'' in \emph{Proc.
  IEEE Wirel. Commun. Netw. Conf.}, 2021, pp. 1--7.

\bibitem{irsembb}
M.~Almekhlafi, M.~A. Arfaoui, M.~Elhattab, C.~Assi, and A.~Ghrayeb, ``Joint
  resource allocation and phase shift optimization for {RIS}-aided {eMBB/URLLC}
  traffic multiplexing,'' \emph{IEEE Trans. Commun.}, pp. 1304 -- 1319, Feb.
  2022.

\bibitem{marirsj}
M.~Najafi, V.~Jamali, R.~Schober, and H.~V. Poor, ``Physics-based modeling and
  scalable optimization of large intelligent reflecting surfaces,'' \emph{IEEE
  Trans. Commun.}, vol.~69, no.~4, pp. 2673--2691, April 2021.

\bibitem{quad}
V.~Jamali, M.~Najafi, R.~Schober, and H.~V. Poor, ``Power efficiency, overhead,
  and complexity tradeoff of {IRS} codebook design—quadratic phase-shift
  profile,'' \emph{IEEE Commun. Letters}, vol.~25, no.~6, pp. 2048--2052, June
  2021.

\bibitem{ghanem7code}
W.~R. Ghanem, V.~Jamali, M.~Schellmann, H.~Cao, J.~Eichinger, and R.~Schober,
  ``Optimization-based phase-shift codebook design for large {IRSs},''
  \emph{arXiv preprint arXiv:2203.01630}, 2022.

\bibitem{Dongfangirsscal}
D.~Xu, V.~Jamali, X.~Yu, D.~W.~K. Ng, and R.~Schober, ``Optimal resource
  allocation design for large irs-assisted {SWIPT} systems: A scalable
  optimization framework,'' \emph{IEEE Trans. Commun.}, vol.~70, no.~2, pp.
  1423--1441, Feb. 2022.

\bibitem{two_time_scaleirs1}
M.-M. Zhao, A.~Liu, Y.~Wan, and R.~Zhang, ``Two-timescale beamforming
  optimization for intelligent reflecting surface aided multiuser communication
  with {QoS} constraints,'' \emph{IEEE Trans. Wirel. Commun.}, vol.~20, no.~9,
  pp. 6179--6194, Sept. 2021.

\bibitem{two_time_scaleirs2}
Y.~Cao, T.~Lv, and W.~Ni, ``Two-timescale optimization for intelligent
  reflecting surface-assisted {MIMO} transmission in fast-changing channels,''
  \emph{IEEE Trans. Wirel. Commun.}, vol. Early Access, pp. 1--1, 2022.

\bibitem{3Gpp2a}
3rd Generation Partnership~Project, ``Technical specification group radio
  access network; evolved universal terrestrial radio access ({E-UTRA});
  physical channels and modulation ({3GPP TS} 36.211 version 12.9.0 release
  12).''

\bibitem{beamirs}
\BIBentryALTinterwordspacing
X.~Wei, L.~Dai, Y.~Zhao, G.~Yu, and X.~Duan, ``Codebook design and beam
  training for extremely large-scale {RIS:} far-field or near-field?'' 2021.
  [Online]. Available: \url{https://arxiv.org/abs/2109.10143}
\BIBentrySTDinterwordspacing

\bibitem{thesis}
Y.~Polyanskiy, ``Channel coding: {N}on-asymptotic fundamental limits,'' Ph.D.
  dissertation, Princeton University.

\bibitem{Leemixed}
J.~Lee and S.~Leyffer, \emph{Mixed Integer Nonlinear Programming}.\hskip 1em
  plus 0.5em minus 0.4em\relax Springer Publishing Company, Incorporated, 2011.

\bibitem{cvx}
M.~Grant and S.~Boyd, ``{CVX}: Matlab software for disciplined convex
  programming, version 2.1,'' \url{http://cvxr.com/cvx}, Mar. 2014.

\bibitem{fastglobal}
H.~H. Kha, H.~D. Tuan, and H.~H. Nguyen, ``Fast global optimal power allocation
  in wireless networks by local {D.C.} programming,'' \emph{{IEEE} Trans.
  Wireless Commun}, vol.~11, no.~2, pp. 510--515, February 2012.

\bibitem{aocon}
P.~Tseng, ``Convergence of a block coordinate descent method for
  nondifferentiable minimization,'' \emph{J. Optim. Theory Appl.}, vol. 109,
  no.~3, pp. 118--125, Jun. 2001.

\bibitem{polik}
\BIBentryALTinterwordspacing
I.~P{\'o}lik and T.~Terlaky, \emph{Interior Point Methods for Nonlinear
  Optimization}.\hskip 1em plus 0.5em minus 0.4em\relax Berlin, Heidelberg:
  Springer Berlin Heidelberg, 2010, pp. 215--276. [Online]. Available:
  \url{https://doi.org/10.1007/978-3-642-11339-0_4}
\BIBentrySTDinterwordspacing

\bibitem{quadriga}
S.~Jaeckel, L.~Raschkowski, K.~B{\"o}rner, and L.~Thiele, ``Quadriga: A 3-d
  multi-cell channel model with time evolution for enabling virtual field
  trials,'' \emph{IEEE Trans. Antennas Propag.}, vol.~62, no.~6, pp.
  3242--3256, June 2014.

\bibitem{3Gpp38901}
3rd Generation Partnership~Project, ``Technical specification group radio
  access network; study on channel model for frequencies from 0.5 to 100 {GHz}
  ({3GPP TR} 38.901 version 16.1.0 (2019-12)).''

\end{thebibliography}
\bibliographystyle{IEEEtran}
\end{document}